\documentclass{myThesisTemp}

\usepackage[outdir=./]{epstopdf}
\title{Voiced-Aware Style Extraction and
Style Direction Adjustment for Expressive Text-to-Speech}{음성 영역 인식을 통한 스타일 추출 및 스타일 방향 조정 기반의 표현력있는 음성 합성}

\author[korean]{김 남 규}{}{}
\author[english]{Kim}{Nam-Gyu}{}

\advisor{이 성 환}{Seong-Whan Lee}

\department{
Artificial Intelligence
}{
인 공 지 능 학 과
}

\committee[1]{Seong-Whan Lee} 
\committee[2]{Won-Zoo Chung}
\committee[3]{Tae-Eui Kam}

\graduateDate{2026}{2}
\submitDate{2025}{11}
\approvalDate{2025}{12}

\captionLineSpacing{150}
\abstractLineSpacing{200}
\krAbstractLineSpacing{200}
\TOCLineSpacing{200}
\contentLineSpacing{200}
\acknowledgementLineSpacing{200}

\begin{document}

\providecommand{\methodname}{}
\renewcommand{\methodname}{Spotlight-TTS}

\providecommand{\weightcomponent}{}
\renewcommand{\weightcomponent}{voiced-aware style extraction}

\chapter{Introduction}

\sloppy
Text-to-speech (TTS) synthesis aims to generate natural-sounding speech from input text~\cite{fastspeech2, tacotron2, vits}. With recent advancements in deep learning~\cite{min2022attentional, lee2020uncertainty, lee1990translation, ahmad2006human, hwang2006full}, the naturalness of synthesized speech has improved significantly~\cite{hierspeech, emospheretts}. Despite the development of general TTS systems, synthesizing human-like speech for applications such as virtual assistants and audiobooks remains challenging due to the lack of expressiveness in synthesized speech. To address this limitation, expressive TTS with style modeling and transfer techniques has attracted increasing attention. In particular, style transfer TTS systems are becoming increasingly important for real-world applications, as they eliminate the need for style-annotated datasets or matched pairs of text and speech data.

A well-designed style encoder is a key component for achieving natural style transfer in TTS systems. Early approaches utilized sentence-level style by applying pooling operations~\cite{gst, styletokens}. However, speaking style cannot be fully expressed by sentence-level representations alone, as they have limitations in capturing temporal variations in speech, which led to syllable-level intonation modeling to better capture time-varying style~\cite{qitts}. Furthermore, recent work explored more fine-grained approaches by extracting frame-level style~\cite{generspeech}. This approach employs multi-level style encoders with vector quantization variational autoencoder (VQ-VAE)~\cite{vqvae} to encode style into discrete codebooks, serving as a bottleneck to eliminate non-style information. This approach has led to improvements in style transfer. However, there remains room for further improvement.

Recent works~\cite{tsptts, tcsinger} have utilized advanced vector quantization methods to improve style extraction, such as residual vector quantization (RVQ) and clustering style encoders. Despite these advances in style extraction methodology, several fundamental challenges remain unsolved. First, these methods treat all temporal segments equally, failing to capture the varying importance of different speech regions in extracting speaking style. Different regions of speech have varying impacts on speaking style, particularly voiced regions, which contain rich acoustic information such as harmonics generated by vocal fold vibration. Second, the straight-through estimator~\cite{ste} used during backpropagation disregards the relative positioning of encoded features within each codebook region, limiting the model's ability to learn fine-grained style details. Finally, relying solely on quantization as a bottleneck without additional constraints does not ensure that extracted style embeddings are independent of content, increasing the risk of content leakage during style transfer. Addressing these issues requires methods that enhance expressiveness while maintaining strong disentanglement from content.

To overcome these limitations, we propose \methodname{}, a novel framework that exclusively emphasizes style via voiced-aware style extraction and style direction adjustment. The importance of region-specific processing has been recognized in signal compression~\cite{signalcompression, lim2000text, lee2001automatic, maeng2012nighttime} and has recently gained renewed attention in deep learning~\cite{maskedvq, goodhelper}. Similarly, different regions of speech have varying impacts on speaking style. Our voiced-aware style extraction enables effective quantization based on a more compact representation, which enhances the preservation of style-related features with finer granularity. Specifically, we focus RVQ processing on voiced frames through voiced extraction, while maintaining continuity across different speech regions using an unvoiced filler module with biased self-attention. Additionally, we replace the conventional straight-through estimator with a rotation trick~\cite{rotationtrick} in our quantization process. This approach, along with advances in feature transformation and geometric methods~\cite{roh2007accurate, yang2007reconstruction}, enables more precise style extraction, particularly in voiced regions. Furthermore, we introduce style direction adjustment, which adjusts the extracted style by modifying its angle using content and prosody vectors in the embedding space. This adjustment process effectively removes content information from style and prevents training instabilities caused by disentanglement.

Our contributions are summarized as follows: 
\begin{itemize}
    \item We introduce voiced-aware style extraction that focuses on style-rich voiced regions while maintaining speech continuity through biased self-attention in the unvoiced filler module. 
    \item We apply the rotation trick to improve gradient flow during quantization, enabling more precise style capture in voiced regions. 
    \item We propose style direction adjustment that disentangles content from style while preserving prosody information, improving both speech quality and pronunciation. 
    \item Experimental results demonstrate that \methodname{} achieves superior performance compared to baseline models in terms of expressiveness, overall speech quality, and style transfer capability.
\end {itemize}

\providecommand{\methodname}{}
\renewcommand{\methodname}{Spotlight-TTS}

\providecommand{\weightcomponent}{}
\renewcommand{\weightcomponent}{voiced-aware style extraction}

\chapter{Background \& Related Work}

\section{Expressive Text-to-Speech}
Text-to-speech (TTS) synthesis has evolved significantly with the advent of deep learning architectures. Early neural TTS systems such as Tacotron~\cite{tacotron} and Tacotron 2~\cite{tacotron2} employed sequence-to-sequence models with attention mechanisms to generate mel-spectrograms from text. WaveNet~\cite{wavenet} introduced autoregressive neural vocoders capable of generating high-quality waveforms. However, these autoregressive models suffered from slow inference speeds. Non-autoregressive models such as FastSpeech~\cite{fastspeech} and FastSpeech 2~\cite{fastspeech2} addressed this limitation by introducing duration predictors and variance adaptors, enabling parallel mel-spectrogram generation. More recently, end-to-end models such as VITS~\cite{vits} have achieved high-quality speech synthesis by directly generating waveforms from text through variational inference and adversarial training.

Despite these advances in naturalness, synthesizing expressive speech that conveys emotions and speaking styles remains challenging. Early expressive TTS approaches focused on emotion-labeled datasets~\cite{esd} and categorical emotion modeling. However, collecting emotion-annotated data is labor-intensive and limited in capturing the continuous nature of emotional expression. This motivated research into style transfer TTS, which extracts style information from reference speech without requiring explicit annotations.


\section{Style Modeling in TTS}

Style modeling is crucial for expressive TTS systems to capture and transfer speaking styles from reference speech. Global Style Tokens (GST)~\cite{gst} introduced a style encoder that uses attention mechanisms over learned style tokens to capture sentence-level prosody. Style Tokens~\cite{styletokens} further developed this approach by using variational autoencoders to model style in a continuous latent space. However, these sentence-level approaches have limitations in representing temporal variations within speech, as they compress the entire utterance into a single vector through pooling operations.

To capture time-varying styles, several approaches have been proposed. Meta-StyleSpeech~\cite{metastylespeech} introduced multi-speaker adaptive TTS with style modeling. QI-TTS~\cite{qitts} explored syllable-level intonation control for emotional speech synthesis. GenerSpeech~\cite{generspeech} took a more fine-grained approach by extracting frame-level style representations using vector quantization variational autoencoder (VQ-VAE)~\cite{vqvae}. This method employs multi-level style encoders with VQ-VAE to encode style into discrete codebooks, which serve as a bottleneck to eliminate non-style information. The discrete representation helps improve style transfer by reducing content leakage.

Recent advances have focused on improving vector quantization for style extraction. TSP-TTS~\cite{tsptts} introduced text-based style prediction with residual vector quantization (RVQ) for expressive TTS. TCSinger~\cite{tcsinger} proposed zero-shot singing voice synthesis with style transfer using clustering style encoders and multi-level style control. Despite these improvements, existing methods treat all temporal segments equally without considering the varying importance of different speech regions for style extraction. Additionally, these methods rely solely on quantization as a bottleneck without explicitly constraining the independence between style and content embeddings.


\section{Vector Quantization}

Vector quantization (VQ) has been widely adopted in representation learning for its ability to learn discrete latent representations. VQ-VAE~\cite{vqvae} introduced a discrete latent representation framework by learning a codebook of vectors and mapping continuous features to the nearest codebook entries. This discretization serves as a strong information bottleneck, which is particularly useful for disentangling different factors of variation in the data.

In TTS, vector quantization has been employed to encode style information into discrete tokens. GenerSpeech~\cite{generspeech} uses VQ-VAE in its style encoder to create a discrete style representation that helps eliminate content information. The quantization process involves encoding continuous features into discrete codes and reconstructing them using codebook embeddings. During training, the straight-through estimator~\cite{ste} is commonly used to approximate gradients through the non-differentiable quantization operation. However, this approach treats the gradient as if the quantization operation were an identity function, which ignores the relative positioning of encoded features within codebook regions.

Recent work has introduced improvements to vector quantization. Residual Vector Quantization (RVQ)~\cite{rvq} proposed a hierarchical quantization scheme where residuals are quantized iteratively across multiple codebook layers, enabling more precise representation of complex features. The rotation trick~\cite{rotationtrick} addressed the limitations of the straight-through estimator by restructuring the quantization operation to preserve gradient information about the relative position of input features with respect to codebook vectors. This approach computes a rotation transformation that aligns the input feature to the nearest codebook vector during the forward pass, while enabling more informative gradients during backpropagation.

In the context of speech synthesis, different acoustic regions contribute unequally to speaking style. Voiced regions, characterized by periodic vibrations of the vocal folds, contain rich harmonic structures that are highly correlated with prosody and emotion. In contrast, unvoiced regions contain aperiodic noise-like signals with simpler spectral patterns. This motivates region-specific processing approaches that have been explored in signal compression~\cite{signalcompression} and recently in deep learning~\cite{maskedvq, goodhelper}. Masked vector quantization~\cite{maskedvq} demonstrated that selectively quantizing important image regions improves autoregressive image generation. Similarly, attention-driven masked image modeling~\cite{goodhelper} showed that focusing on informative regions enhances representation learning. These insights suggest that region-aware quantization could be beneficial for style extraction in speech synthesis.
\providecommand{\methodname}{}
\renewcommand{\methodname}{Spotlight-TTS}

\providecommand{\weightcomponent}{}
\renewcommand{\weightcomponent}{voiced-aware style extraction}

\chapter{Methods}

\begin{figure}[!t]
\centering
\includegraphics[width=0.95\textwidth]{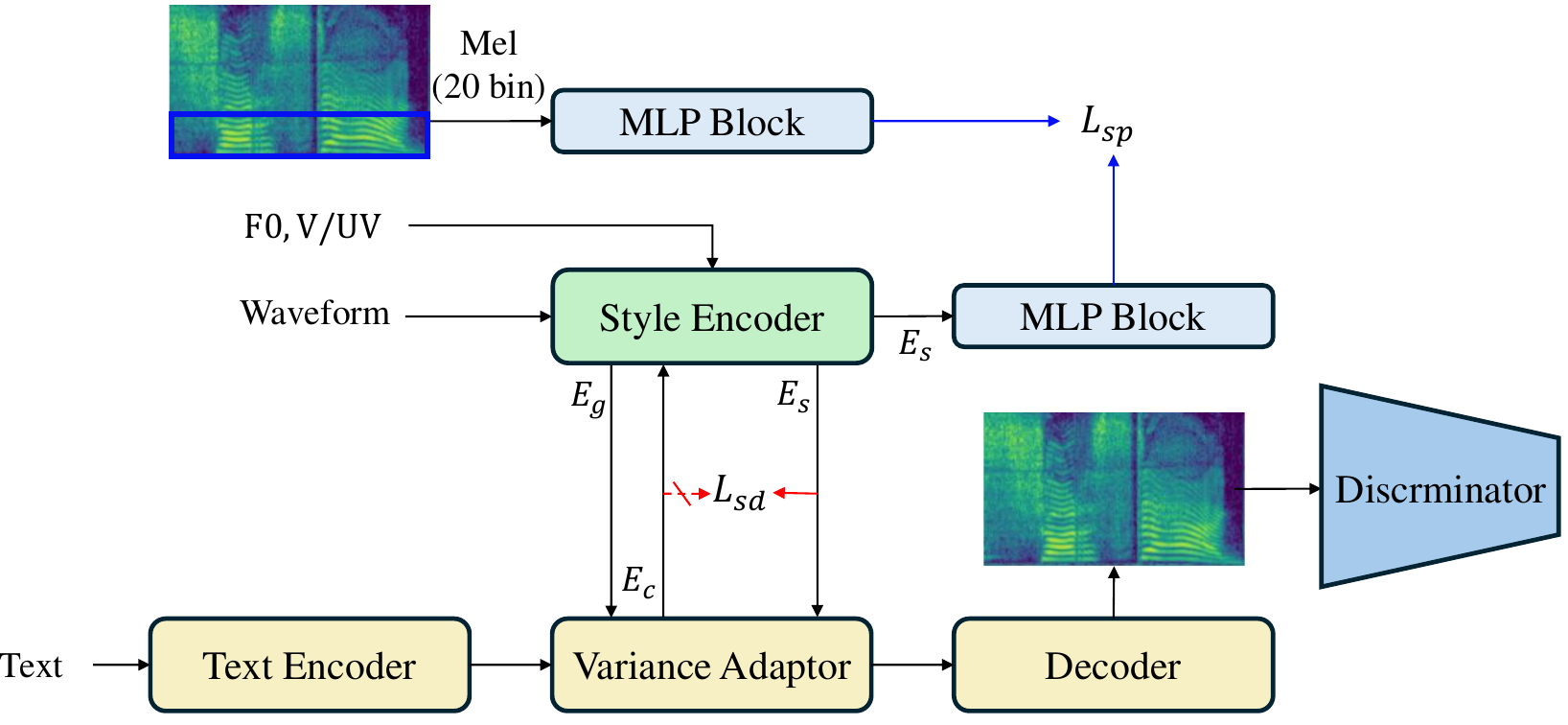}
\caption{Overall architecture of Spotlight-TTS. $E_c$, $E_g$, and $E_s$ denote the content embedding, global style embedding, and style embedding respectively.}
\label{fig:overall_architecture}
\end{figure}

In this chapter, we introduce the key methodologies of \methodname{} that form the foundation of our approach for expressive text-to-speech synthesis. As illustrated in Figure~\ref{fig:overall_architecture}, our model consists of a text encoder, variance adaptor, decoder, discriminators, style encoder, and MLP blocks. The overall architecture follows FastSpeech 2~\cite{fastspeech2} with multi-level discriminators~\cite{syntaspeech} for improved speech quality.

Our proposed method focuses on two main contributions. First, we present voiced-aware style extraction, which focuses on voiced regions that are highly correlated with speaking style while maintaining continuity across different speech regions through an unvoiced filler module. This approach enables effective quantization based on a more compact representation, enhancing the preservation of style-related features with finer granularity. Second, we introduce style direction adjustment, which adjusts the extracted style by modifying its angle using content and prosody vectors in the embedding space. This adjustment process effectively removes content information from style and prevents training instabilities caused by disentanglement.

The style encoder in our model consists of two components: a pre-trained global style encoder that extracts sentence-level style, and a voiced-aware style encoder that focuses on frame-level style. The global style encoder follows the architecture from GenerSpeech~\cite{generspeech}, while the voiced-aware style encoder is composed of a WaveNet block, four convolutional residual blocks, a residual vector quantization (RVQ) module with rotation trick, and three unvoiced filler blocks.

\begin{figure}[!t]
\centering
\includegraphics[width=0.85\textwidth]{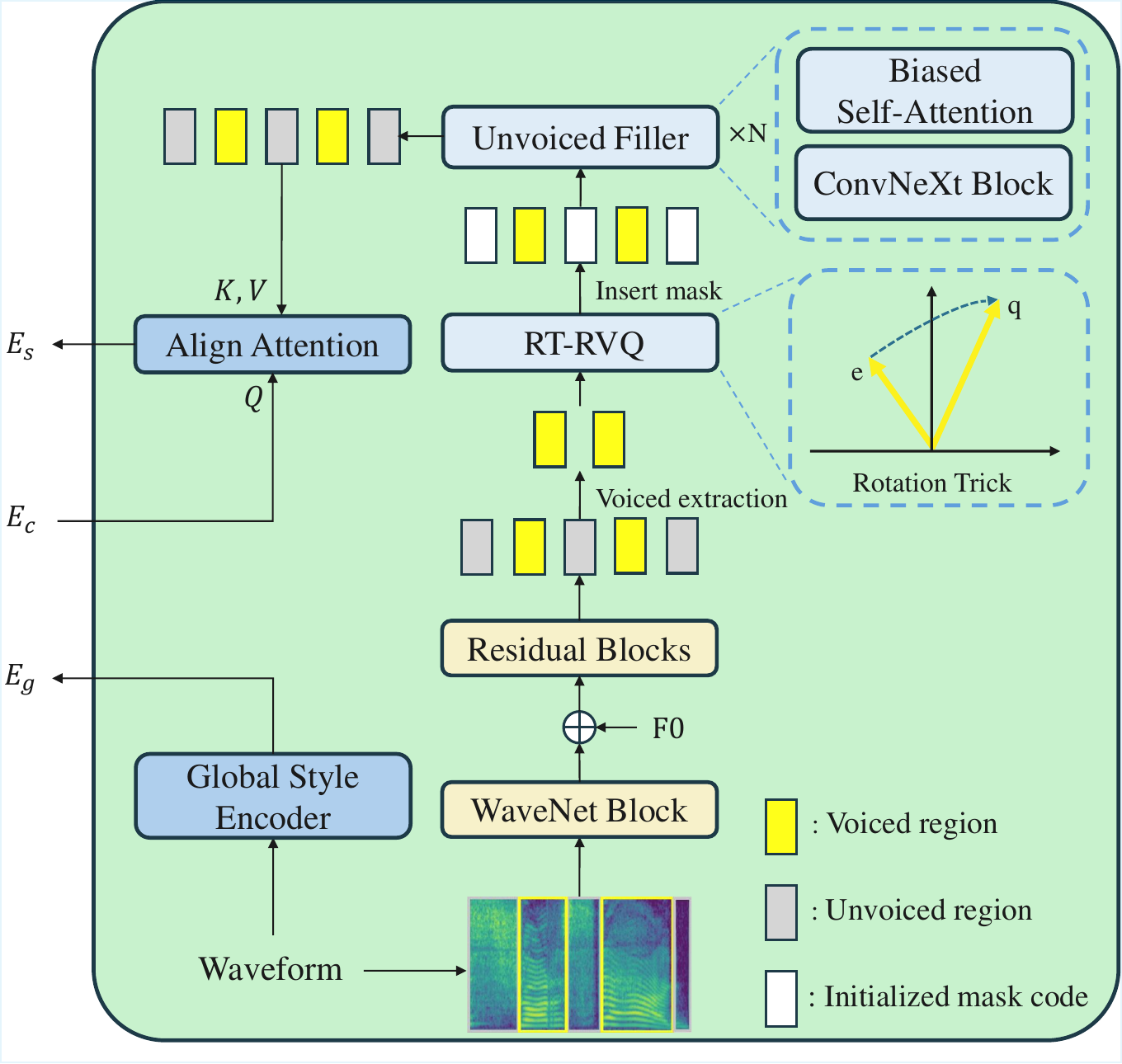}
\caption{Detailed architecture of the style encoder. The encoder processes voiced regions through RT-RVQ (Rotation Trick with Residual Vector Quantization) and fills unvoiced regions using the Unvoiced Filler module with biased self-attention. Yellow regions indicate voiced frames, gray regions indicate unvoiced frames, and dotted boxes represent initialized mask code embeddings. $e$ and $q$ represent the input feature and quantized vector respectively.}
\label{fig:style_encoder}
\end{figure}

\section{Voiced-aware Style Extraction}

In speech synthesis, we hypothesize that style extraction through codebook learning can be improved by considering different mel-spectrogram regions that contribute unequally to speaking style. As shown in Figure~\ref{fig:style_encoder}, voiced regions consist of harmonics that are highly correlated with speaking style, and unvoiced regions have simple repetitive patterns that are less related to style. Different regions of speech have varying impacts on speaking style. Voiced regions, which contain rich acoustic information such as harmonics generated by vocal fold vibration, are particularly important for style characteristics. Therefore, we propose voiced-aware style extraction, a novel style extraction method that focuses on the voiced region and fills in the unvoiced region with the unvoiced filler module.

\subsection{Voiced Frame Processing}

We use an RVQ module~\cite{rvq} to extract detailed style embeddings. Given that unvoiced regions are less relevant to style compared to voiced regions, we focus RVQ processing on voiced frames through voiced extraction (VE). We use pre-extracted voiced and unvoiced (V/UV) flags to aggregate only voiced frames from the intermediate features as input to the RVQ module.

During quantization, we adopt the rotation trick (RT)~\cite{rotationtrick} to improve gradient flow through the quantization layer. Unlike the conventional straight-through estimator~\cite{ste}, RT preserves the angle between the loss gradient and codebook vector during backpropagation. The RT is computed as:
\begin{equation}
\tilde{q} = \text{sg}\left[\frac{\|q\|}{\|e\|} R\right] e,
\end{equation}
where $R$ is the rotation transformation that aligns the input feature $e$ to the closest codebook vector $q$, and $\frac{\|q\|}{\|e\|}$ rescales $e$ to match the magnitude of $q$. Here, $\text{sg}[\cdot]$ denotes the stop-gradient operator used to detach rotation and scaling terms from gradient computation.

During the forward pass, we use $\tilde{q}$, the transformed input feature, instead of the quantized vector. $\tilde{q}$ is identical to $q$ in the forward pass, ensuring the RVQ output remains unchanged. By rewriting $q$ in terms of $e$, the backward pass rotates the gradient to capture the relative position of $e$ in the codebook. This enables the model to learn more precise style representations, particularly in voiced regions where harmonic structures are crucial for speaking style.

The RVQ module has a depth of four, applying residual quantization iteratively across multiple codebook layers. This hierarchical quantization scheme enables more precise representation of complex features. The RT is computed using a Householder reflection matrix, which provides an efficient and stable rotation transformation.

\subsection{Unvoiced Filler Module}

To improve style continuity between different regions, we propose an unvoiced filler (UF) module. After quantization, we generate learnable mask code embeddings with uniform random initialization and insert them into the unvoiced positions based on their positional information. The UF module then fills these embeddings with meaningful acoustic information.

The UF module consists of $N$ identical sub-modules, each of which consists of a ConvNeXt block~\cite{convnext} and a biased self-attention mechanism. Biased self-attention enables information flow from the non-masked regions to mask code regions while blocking the opposite direction. Through this asymmetric information flow, the model can utilize the non-masked region without the mask code region negatively affecting the non-masked region.

The biased self-attention is defined as follows:
\begin{equation}
\text{Attention}(q, k, v) = \text{SoftMax}\left(\frac{qk^T}{\sqrt{d}} \odot \beta\right) v,
\end{equation}
where $\beta$ is the attention reweighting (AR) coefficient. Specifically, we define $\beta$ as $0.02$ for mask positions and $1$ for non-masked positions to achieve asymmetric information flow.

This design allows the model to maintain natural prosodic continuity between voiced and unvoiced regions. The small but non-zero coefficient ($0.02$) for mask positions prevents complete information blocking while still emphasizing the directional flow from voiced to unvoiced regions. This is crucial because while unvoiced regions are less important for style extraction, they still require acoustic information to maintain natural speech flow.

After the unvoiced filler blocks, we use scaled dot-product attention~\cite{attention} to align the time dimension of the style embedding with that of the content embedding. The aligned style embedding is utilized within the variance adaptor and added to its final output along with the global style embedding.

\section{Style Direction Adjustment}

\begin{figure}[!t]
\centering
\includegraphics[width=0.7\textwidth]{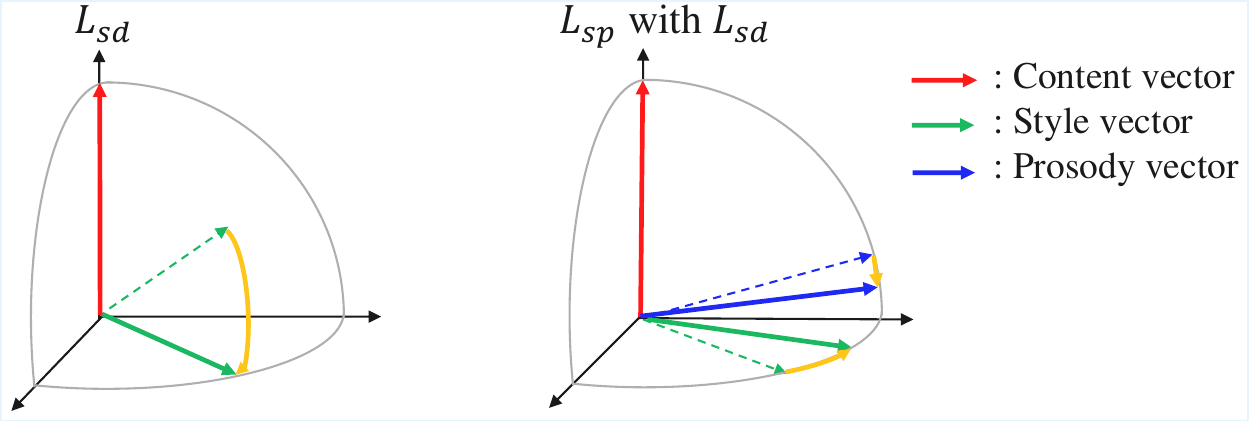}
\caption{Conceptual illustration of style direction adjustment. The figure shows how the direction of the style vector (orange) is adjusted in the embedding space. The style disentanglement loss $\mathcal{L}_{sd}$ encourages orthogonality between the style vector and content vector (blue), while the style preserving loss $\mathcal{L}_{sp}$ maintains alignment with the prosody vector (green).}
\label{fig:style_direction}
\end{figure}

After extracting the style, we adjust the directionality of the style to remove content information. As conceptually illustrated in Figure~\ref{fig:style_direction}, we introduce two complementary losses: style disentanglement (SD) loss encouraging orthogonality between style and content, and style preserving (SP) loss to align the style with prosody. The left side shows the initial state where the style vector contains content information, and the right side shows the adjusted state after applying both losses.

\subsection{Style Disentanglement Loss}

Inspired by cross-speaker emotion disentangling~\cite{emotiondisentangle}, we use orthogonality loss to disentangle style and content information. Since content information within style embedding can interfere with the learning of content embedding, we detach the content embedding during training. The SD loss $\mathcal{L}_{sd}$ is defined as follows:
\begin{equation}
\mathcal{L}_{sd} = \left\|\text{sg}[E_c] E_s^T\right\|_F^2,
\end{equation}
where $E_c$ and $E_s$ denote the content embedding and style embedding, respectively. The expression $\|\cdot\|_F^2$ is the squared Frobenius norm.

This orthogonality constraint encourages the style embedding to be independent of content information. By minimizing the inner product between content and style embeddings, we ensure that style representations do not contain linguistic information that could leak during style transfer. The stop-gradient operation on the content embedding ensures that this constraint primarily affects the style encoder rather than interfering with content learning.

\subsection{Style Preserving Loss}

To enhance stability in style learning and mitigate errors from F0 and V/UV predictions, we introduce a style preserving (SP) loss. This loss refines extracted style embeddings by increasing similarity with low-frequency prosody embeddings. Two separate MLP blocks are used to extract prosody embeddings from the lower 20 bins of the mel-spectrogram and style embedding, respectively.

By increasing the cosine similarity between these two embeddings, we refine the prosody information in the style embedding. The SP loss $\mathcal{L}_{sp}$ is defined as follows:
\begin{equation}
\mathcal{L}_{sp} = -\sum_{i=1}^{t} \cos\_\text{sim}(p_i, \tilde{s}_i),
\end{equation}
where $t$ is the number of timesteps, $p_i$ represents the prosody vector extracted from the low-band mel-spectrogram, and $\tilde{s}_i$ denotes the projected style vector.

Each MLP block consists of two linear layers with a GELU activation function~\cite{gelu}. For the SP loss, both style embedding and the lower 20 bins of mel-spectrogram are projected to 32 dimensions through MLP blocks from their initial dimensions of 256 and 20, respectively.

The low-frequency region of the mel-spectrogram (lower 20 bins) primarily contains fundamental frequency and low-order harmonic information, which are strongly correlated with prosody. By aligning the style embedding with this prosodic information, we ensure that the style representation preserves essential prosodic characteristics even after the disentanglement process. This complementary loss prevents the SD loss from being too aggressive and removing prosody-related information along with content.

\section{Training Objective}

The proposed \methodname{} model is trained using multiple objective functions. The overall loss function is formulated as follows:
\begin{equation}
\mathcal{L}_{\text{total}} = \mathcal{L}_{\text{fs2}} + \lambda_{\text{rvq}}\mathcal{L}_{\text{rvq}} + \lambda_{\text{adv}}\mathcal{L}_{\text{adv}} + \lambda_{\text{sd}}\mathcal{L}_{sd} + \lambda_{\text{sp}}\mathcal{L}_{sp},
\end{equation}
where $\mathcal{L}_{\text{fs2}}$ represents the FastSpeech 2 loss~\cite{fastspeech2}, which includes mel-spectrogram reconstruction loss, duration loss, pitch loss, and energy loss. $\mathcal{L}_{\text{rvq}}$ is the RVQ loss for training the vector quantization module, computed as the sum of commitment loss and codebook loss across all quantization layers. $\mathcal{L}_{\text{adv}}$ denotes the adversarial loss from the multi-length discriminator~\cite{syntaspeech}, which improves speech quality through adversarial training. $\mathcal{L}_{sd}$ and $\mathcal{L}_{sp}$ are the style disentanglement loss and style preserving loss, respectively.

The hyperparameters $\lambda_{\text{rvq}}$, $\lambda_{\text{adv}}$, $\lambda_{\text{sd}}$, and $\lambda_{\text{sp}}$ control the relative importance of each loss term. In our experiments, we set these values to $1.0$, $0.05$, $0.02$, and $0.02$, respectively. These values were determined through preliminary experiments to balance speech quality, style transfer performance, and training stability.

By minimizing this overall loss function, the model is trained to not only generate high-quality mel-spectrograms but also extract expressive style representations that are disentangled from content while preserving prosodic information. The voiced-aware style extraction focuses on style-rich regions, while the style direction adjustment ensures proper separation between content and style in the embedding space.
\providecommand{\methodname}{}
\renewcommand{\methodname}{Spotlight-TTS}

\providecommand{\weightcomponent}{}
\renewcommand{\weightcomponent}{voiced-aware style extraction}

\chapter{Experiments}

\section{Experimental Setup}
\sloppy
We use the Emotional Speech Dataset (ESD)~\cite{esd} to verify whether the models can capture style using expressive reference speech. The ESD contains ten English speakers, each producing 350 sentences in five emotions (happy, sad, neutral, surprise, and angry). We follow the original partitioning criteria of the dataset, totaling 17,500 samples. The training, validation, and test sets contain 14,000, 1,750, and 1,750 samples, respectively.

Mel-spectrograms with 80 bins are extracted using short-time Fourier transform with a hop size of 256, a window size of 1,024, and an FFT size of 1,024. For the audio synthesis in our experiments, we train the vocoder using the official BigVGAN implementation~\cite{bigvgan}. 

Our TTS system is based on FastSpeech 2~\cite{fastspeech2} and incorporates a multi-length discriminator~\cite{syntaspeech} for improved speech quality. The style encoder is composed of two parts: a pre-trained global style encoder that extracts the sentence-level style and a voiced-aware style encoder that focuses on frame-level style. The global style encoder follows the architecture from GenerSpeech~\cite{generspeech}. The voiced-aware style encoder consists of a WaveNet block, four convolutional residual blocks, an RVQ module, and three unvoiced filler (UF) blocks. The RVQ module has a depth of four and applies rotation trick computed by a Householder reflection matrix.

For style preserving (SP) loss, both style embedding and lower 20 bins of mel-spectrogram are projected to 32 dimensions through MLP blocks from their initial dimensions of 256 and 20, respectively. Each MLP block consists of two linear layers with a GELU activation function~\cite{gelu}. After the unvoiced filler blocks, we use scaled dot-product attention~\cite{attention} to align the time dimension of the style embedding with that of the content embedding.

We employ the AdamW optimizer~\cite{adamw}, setting the hyperparameters $\beta_1$ to 0.9 and $\beta_2$ to 0.98. Our model is trained for 200,000 steps on a single NVIDIA RTX 2080Ti GPU. For the loss weights, we set $\lambda_{\text{rvq}}$, $\lambda_{\text{adv}}$, $\lambda_{\text{sd}}$, and $\lambda_{\text{sp}}$ to $1.0$, $0.05$, $0.02$, and $0.02$, respectively.

We compare our \methodname{} with other FastSpeech2-based style transfer TTS models: FastSpeech2-GST~\cite{fastspeech2, gst}, FastSpeech2-CSE~\cite{fastspeech2, tcsinger}, StyleSpeech~\cite{metastylespeech}, and GenerSpeech~\cite{generspeech}. For models utilizing time-variant style, we use the same global style encoder for sentence-level style. For reference audio selection during inference, we follow the same strategy as in GenerSpeech.

\subsection{Evaluation Metrics}

To evaluate the speech quality, we conduct both objective and subjective evaluations of the synthesized speech. For subjective evaluation, we use the naturalness mean opinion score (nMOS) and similarity mean opinion score (sMOS). Both metrics are rated from 1 to 5 with a confidence interval of 95\%. We randomly select 50 samples from the test set, with 5 samples per speaker. Each audio sample is listened to by 20 participants.

Additionally, we utilize the open-source UTMOS~\cite{utmos} as a MOS prediction model for the naturalness metric. To evaluate linguistic consistency, we calculate the word error rate (WER) using Whisper~\cite{whisper}. For speaker similarity measurements, we calculate the speaker embedding cosine similarity (SECS) via WavLM~\cite{wavlm}.

For prosodic evaluation, we compute the root mean square error for both pitch error ($\text{RMSE}_{f0}$) measured in Hz and periodicity error ($\text{RMSE}_p$), along with the F1 score of voiced/unvoiced classification ($\text{F1}_{v/uv}$). All objective metrics are computed using the entire test set.

\section{Main Results}

Table~\ref{tab:main_results} presents a comprehensive comparison of \methodname{} with baseline models across subjective and objective metrics. Our proposed model achieves an nMOS of $4.26 \pm 0.04$, which is the highest among all TTS models and approaches the quality of BigVGAN ($4.30 \pm 0.04$). This represents a significant improvement over the best baseline model (GenerSpeech: $3.98 \pm 0.04$), with an improvement of approximately 0.28 points.

\begin{table*}[!t]
\renewcommand{\arraystretch}{1.1}
\caption{Comparison with different models for subjective and objective metrics. Style Variation indicates whether the model supports time-variant style modeling. Higher values are better for all metrics except WER, RMSE$_{f0}$, and RMSE$_p$, where lower values indicate better performance.}
\centering
\resizebox{\textwidth}{!}{%
\begin{tabular}{l|c|cc|cccccc}
\hline
\multirow{2}{*}{Method} & Style & \multicolumn{2}{c|}{Subjective} & \multicolumn{6}{c}{Objective} \\ \cline{3-10}
 & Variation & nMOS ($\uparrow$) & sMOS ($\uparrow$) & UTMOS ($\uparrow$) & WER ($\downarrow$) & RMSE$_{f0}$ ($\downarrow$) & RMSE$_p$ ($\downarrow$) & F1$_{v/uv}$ ($\uparrow$) & SECS ($\uparrow$) \\ \hline
GT & - & $4.34 \pm 0.04$ & $4.63 \pm 0.02$ & 3.78 & 12.42 & - & - & - & 0.9168 \\
BigVGAN & - & $4.30 \pm 0.04$ & $4.58 \pm 0.03$ & 3.63 & 12.40 & 2.45 & 0.2749 & 0.8008 & 0.9150 \\ \hline
FastSpeech2 w/ GST & $\times$ & $3.77 \pm 0.05$ & $3.05 \pm 0.04$ & 3.39 & 14.18 & 13.37 & 0.4619 & 0.6707 & 0.8945 \\
FastSpeech2 w/ CSE & \checkmark & $3.74 \pm 0.05$ & $3.46 \pm 0.04$ & 3.42 & 13.49 & 10.39 & 0.4159 & 0.7024 & 0.9013 \\
StyleSpeech & $\times$ & $3.74 \pm 0.05$ & $3.14 \pm 0.04$ & 3.37 & 13.24 & 13.88 & 0.4433 & 0.6716 & 0.9008 \\
GenerSpeech & \checkmark & $3.98 \pm 0.04$ & $3.37 \pm 0.04$ & 2.92 & 16.45 & 11.20 & 0.4343 & 0.6709 & 0.8848 \\ \hline
\methodname{} (Proposed) & \checkmark & $\mathbf{4.26 \pm 0.04}$ & $\mathbf{3.84 \pm 0.04}$ & $\mathbf{3.56}$ & $\mathbf{12.64}$ & $\mathbf{8.27}$ & $\mathbf{0.4050}$ & $\mathbf{0.7053}$ & $\mathbf{0.9061}$ \\ \hline
\end{tabular}%
}
\label{tab:main_results}
\end{table*}

For style similarity, \methodname{} achieves an sMOS of $3.84 \pm 0.04$, substantially outperforming all baseline models. This represents an improvement of 0.47 points over GenerSpeech ($3.37 \pm 0.04$) and demonstrates the effectiveness of our voiced-aware style extraction approach in capturing expressive speaking styles.

In terms of objective metrics, our model achieves the best UTMOS score of 3.56, indicating high predicted naturalness. The WER of 12.64 is the lowest among all TTS models and is comparable to BigVGAN (12.40), demonstrating excellent pronunciation clarity. This improvement over GenerSpeech (WER: 16.45) highlights the effectiveness of our style direction adjustment in preventing content leakage.

For prosodic metrics, \methodname{} demonstrates superior performance across all measures. The pitch error ($\text{RMSE}_{f0}$) of 8.27 Hz is significantly lower than all baselines, with an improvement of 2.12 Hz over FastSpeech2 w/ CSE (10.39 Hz). The periodicity error ($\text{RMSE}_p$) of 0.4050 and F1 score of voiced/unvoiced classification ($\text{F1}_{v/uv}$) of 0.7053 are also the best among all models. These results demonstrate that our voiced-aware style extraction effectively captures prosodic information while maintaining accurate pitch and voicing characteristics.

The speaker embedding cosine similarity (SECS) of 0.9061 indicates excellent speaker similarity, outperforming most baseline models except for FastSpeech2 w/ CSE (0.9013). This demonstrates that our style direction adjustment effectively disentangles content from style without compromising speaker identity.

Global style-based models like FastSpeech2 w/ GST and StyleSpeech demonstrate lower performance in style-related metrics due to the loss of temporal style variations through pooling operations. In contrast, FastSpeech2 w/ CSE and GenerSpeech outperform global style-based models across all style-related metrics by preserving temporal dynamics. Furthermore, our model surpasses all baselines across all metrics by considering temporal dynamics with region-specific importance and directionality of extracted style.

\section{Style Transfer Evaluation}

We evaluate style transfer performance through an AXY preference test with scores ranging from $-3$ to $3$, where $0$ denotes "Both are about the same distance". In this test, X represents the baseline model, Neutral represents equal preference, and Y represents our \methodname{}. Table~\ref{tab:axy_test} shows the results for both parallel and non-parallel style transfer settings.

\begin{table*}[!t]
\renewcommand{\arraystretch}{1.1}
\caption{Results of AXY preference test on parallel and non-parallel style transfer. Higher scores and Y preference percentages indicate better style transfer performance of \methodname{}.}
\centering
\resizebox{0.85\textwidth}{!}{%
\begin{tabular}{l|l|c|ccc}
\hline
\multirow{2}{*}{Baseline} & \multirow{2}{*}{Setting} & 7-point & \multicolumn{3}{c}{Preference (\%)} \\ \cline{4-6}
 &  & score & X & Neutral & Y \\ \hline
\multirow{2}{*}{FastSpeech2 w/ GST} & Parallel & $1.21 \pm 0.12$ & 15\% & 22\% & 63\% \\
 & Non-Parallel & $0.59 \pm 0.11$ & 28\% & 16\% & 56\% \\ \hline
\multirow{2}{*}{FastSpeech2 w/ CSE} & Parallel & $0.53 \pm 0.12$ & 28\% & 17\% & 55\% \\
 & Non-Parallel & $0.33 \pm 0.15$ & 29\% & 29\% & 42\% \\ \hline
\multirow{2}{*}{StyleSpeech} & Parallel & $1.16 \pm 0.11$ & 17\% & 21\% & 62\% \\
 & Non-Parallel & $0.25 \pm 0.12$ & 34\% & 22\% & 44\% \\ \hline
\multirow{2}{*}{GenerSpeech} & Parallel & $0.95 \pm 0.12$ & 17\% & 36\% & 47\% \\
 & Non-Parallel & $0.38 \pm 0.13$ & 24\% & 33\% & 43\% \\ \hline
\end{tabular}%
}
\label{tab:axy_test}
\end{table*}

Our subjective evaluation demonstrates superior style transfer performance across both parallel and non-parallel settings compared to baseline models. In parallel settings, where the reference speech and input text match, \methodname{} achieves the highest preference scores against all baselines. Against FastSpeech2 w/ GST, our model achieves a score of $1.21 \pm 0.12$ with 63\% preference, demonstrating significant improvement in capturing and transferring expressive styles.

Similar to the main results, time-variant style-based models outperform global style-based models in style transfer tasks. \methodname{} further advances this approach by explicitly modeling the relative importance of different temporal regions based on their acoustic characteristics. Unlike previous time-variant style-based models that treated all temporal regions equally, our region-aware style extraction enables more effective style capture, leading to superior style transfer performance.

In non-parallel settings, where the linguistic content differs between reference speech and input text, our model maintains competitive performance. The score of $0.59 \pm 0.11$ against FastSpeech2 w/ GST with 56\% preference demonstrates robust style transfer capability even with content mismatch. Additionally, our style direction adjustment disentangles content information from style in the embedding space, enabling robust style extraction even when linguistic content differs between reference speech and input text.

The performance gap between parallel and non-parallel settings is smaller for \methodname{} compared to baseline models, indicating better generalization in challenging style transfer scenarios. This demonstrates that our voiced-aware style extraction and style direction adjustment work synergistically to maintain style transfer quality across different content conditions.

\section{Ablation Study}

\subsection{Voiced-aware Style Extraction}

We evaluate voiced-aware style extraction by removing key components and analyzing their effects. Table~\ref{tab:ablation_extraction} presents the results of this ablation study.

\begin{table*}[!t]
\renewcommand{\arraystretch}{1.1}
\caption{Results of ablation study on voiced-aware style extraction and style direction adjustment. RT: Rotation Trick, UF: Unvoiced Filler, VE: Voiced Extraction, SP: Style Preserving loss, SD: Style Disentanglement loss.}
\centering
\resizebox{0.75\textwidth}{!}{%
\begin{tabular}{l|ccccc}
\hline
Method & nMOS & WER & RMSE$_{f0}$ & RMSE$_p$ & F1$_{v/uv}$ \\ \hline
Ours (Full) & $3.93 \pm 0.07$ & 12.64 & 8.27 & 0.4050 & 0.7053 \\ \hline
$-$ RT & $3.91 \pm 0.06$ & 13.24 & 9.43 & 0.4154 & 0.6928 \\
$-$ RT $-$ UF & $3.91 \pm 0.07$ & 13.41 & 9.82 & 0.4274 & 0.6874 \\
$-$ RT $-$ UF $-$ VE & $3.84 \pm 0.07$ & 14.06 & 11.48 & 0.4425 & 0.6829 \\ \hline
$-$ SP & $3.86 \pm 0.06$ & 13.66 & 9.74 & 0.4297 & 0.6915 \\
$-$ SP $-$ SD & $3.66 \pm 0.07$ & 15.38 & 8.53 & 0.4037 & 0.6848 \\ \hline
\end{tabular}%
}
\label{tab:ablation_extraction}
\end{table*}

Removing the rotation trick (RT) leads to degraded performance, particularly in style-related metrics. The pitch error increases from 8.27 Hz to 9.43 Hz, and the F1 score for voiced/unvoiced classification decreases from 0.7053 to 0.6928. This indicates that the rotation trick allows the RVQ to better capture the harmonic structures in voiced regions by ensuring stable gradient propagation.

When both the RT and the unvoiced filler (UF) are removed, we observe further performance degradation. The pitch error increases to 9.82 Hz, and the WER increases from 13.24 to 13.41. Although unvoiced regions have relatively simple and repetitive patterns compared to voiced regions, removing the UF module leads to degraded pronunciation and prosody. This indicates that while unvoiced regions are less critical for style, proper handling of these regions through the UF module is still necessary.

Additionally, removing voiced extraction (VE) results in significantly increased pitch error (11.48 Hz) and WER (14.06). This demonstrates that our voiced-aware approach, which aggregates voiced frames to focus quantization on style-rich regions, is essential for capturing detailed style information. The F1 score for voiced/unvoiced classification also decreases to 0.6829, indicating degraded voicing detection.

\subsection{Style Direction Adjustment}

We also investigate the effectiveness of our style direction adjustment mechanism. When the style preserving (SP) loss is removed, we observe highly increased pitch errors (9.74 Hz). This suggests that SP loss effectively mitigates the degradation of prosodic information caused by the strong constraints of style disentanglement (SD) loss.

Removing both SP and SD losses results in severely degraded nMOS (3.66) and WER (15.38), indicating that removing content information from style is important for both speech quality and pronunciation. The increased WER demonstrates that without style direction adjustment, content information leaks into the style embedding, causing pronunciation errors during style transfer.

Interestingly, the pitch error when both losses are removed (8.53 Hz) is lower than when only SP is removed (9.74 Hz). This suggests that SD loss, while necessary for disentanglement, can negatively impact prosody without the balancing effect of SP loss. The combination of both losses achieves the best balance between content disentanglement and prosody preservation.

\subsection{Biased Self-Attention}

To investigate the effectiveness of biased self-attention in the unvoiced filler module, we conduct an additional ablation study presented in Table~\ref{tab:ablation_attention}.

\begin{table}[!t]
\renewcommand{\arraystretch}{1.1}
\caption{Results of ablation study on biased self-attention in unvoiced filler. BM: Binary Mask.}
\centering
\resizebox{0.65\textwidth}{!}{%
\begin{tabular}{l|ccccc}
\hline
Method & nMOS & WER & RMSE$_{f0}$ & RMSE$_p$ & F1$_{v/uv}$ \\ \hline
Ours (Biased Attention) & $3.95 \pm 0.07$ & 12.64 & 8.27 & 0.4050 & 0.7053 \\ \hline
Self-attention w/ BM & $3.93 \pm 0.07$ & 13.61 & 13.19 & 0.4528 & 0.6751 \\
Self-attention & $3.87 \pm 0.07$ & 13.64 & 16.38 & 0.4587 & 0.6668 \\ \hline
\end{tabular}%
}
\label{tab:ablation_attention}
\end{table}

If we replace the attention reweighting (AR) with a simple binary mask (BM) with values of 1 and 0, the model completely blocks information flow from non-masked to mask code regions. This prevents the mask code embeddings from being filled with meaningful acoustic information, disrupting the natural prosodic continuity between non-masked and mask code regions. The result is higher F0 errors (13.19 Hz) and worse V/UV classification (0.6751).

In contrast, using conventional self-attention allows excessive interference between regions, significantly degrading both metrics. The pitch error increases dramatically to 16.38 Hz, and the F1 score decreases to 0.6668. These results validate that our biased self-attention enables optimal information flow between voiced and unvoiced regions, maintaining natural prosody while preserving the benefits of region-specific processing.
\providecommand{\methodname}{}
\renewcommand{\methodname}{Spotlight-TTS}

\providecommand{\weightcomponent}{}
\renewcommand{\weightcomponent}{voiced-aware style extraction}

\chapter{Conclusion}

We presented \methodname{}, a framework for synthesizing expressive speech by focusing on voiced regions in the mel-spectrogram and adjusting the direction of the extracted style. Voiced-aware style extraction considers the acoustic characteristics of different speech regions, enabling more detailed style extraction. Furthermore, the style direction adjustment effectively disentangles content from style, while preserving prosody information within the style embedding.

In conclusion, \methodname{} effectively integrates voiced-aware style extraction and style direction adjustment to enhance expressiveness and speech quality. Our empirical results demonstrate that \methodname{} outperforms baseline models in terms of naturalness, style similarity, and pronunciation accuracy across various evaluation metrics, showcasing its effectiveness for expressive text-to-speech synthesis.

\end{document}